\documentclass[aip, jcp, amsmath, amssymb, reprint]{revtex4-1}
\usepackage{graphicx}
\usepackage{dcolumn}% Align table columns on decimal point
\usepackage{bm}% bold math
\usepackage{color}
\usepackage{natbib}
\begin{document}

\author{John Condoluci}
\affiliation{School of Chemistry and Biochemistry, Georgia Institute of Technology, Atlanta, GA 30332, USA}

\author{Smitha Janardan}
\affiliation{School of Chemistry and Biochemistry, Georgia Institute of Technology, Atlanta, GA 30332, USA}

\author{Aaron T. Calvin}
\affiliation{School of Chemistry and Biochemistry, Georgia Institute of Technology, Atlanta, GA 30332, USA}

\author{Ren\'e Rugango}
\affiliation{School of Chemistry and Biochemistry, Georgia Institute of Technology, Atlanta, GA 30332, USA} 

\author{Gang Shu}
\affiliation{School of Chemistry and Biochemistry, Georgia Institute of Technology, Atlanta, GA 30332, USA}

\author{Kenneth R. Brown}
\email{kenbrown@gatech.edu}
\affiliation{School of Chemistry and Biochemistry, Georgia Institute of Technology, Atlanta, GA 30332, USA}
\affiliation{School of Computational Science and Engineering, Georgia Institute of Technology, Atlanta, GA 30332, USA}
\affiliation{School of Physics, Georgia Institute of Technology, Atlanta, GA 30332, USA}

\title{Reassigning the CaH$^+$ 1$^{1}\Sigma\rightarrow$ 2$^{1}\Sigma$ vibronic transition with CaD$^+$}

\date{\today}

%\citenum{bibid} will put an in-line citation number

\begin{abstract}
We observe vibronic transitions in CaD$^+$  between the 1$^{1}\Sigma$ and 2$^{1}\Sigma$ electronic states by resonance enhanced multiphoton photodissociation spectroscopy in a Coulomb crystal. The vibronic transitions are compared with previous measurements on CaH$^+$. The result is a revised assignment of the CaH$^+$ vibronic levels and a disagreement with CASPT2 theoretical calculations by approximately 700 cm$^{-1}$. 

\end{abstract}

\pacs{}% insert suggested PACS numbers in braces on next line

\maketitle %\maketitle must follow title, authors, abstract and \pacs

\section{\label{sec:level1}Introduction}

Co-trapping molecular ions with Doppler-cooled atomic ions sympathetically cools the molecular motion to millikelvin temperatures, enabling studies in spectroscopy and reaction dynamics. \cite{Larson86, Drewsen2000, Koelemeij2007, Hudson2016, Willitsch2008} Cold ionic ensembles in a variety of platforms have proved useful for astrochemical identification \cite{Campbell2015, Staanum2008, Brunken2014a} and studies of internal state distributions,\cite{Koelemeij2007, Endres2016} while co-trapping with Doppler-cooled ions offers advantages for probing the possible time variation of fundamental constants, \cite{Kajita2011, Bressel2012, Blythe2005} and performing quantum logic spectroscopy. \cite{SchmidtScience2005, Hume2011, Wolf2016, Chou2017} Ionic metal-hydrides like CaH$^+$  and MgH$^+$ are promising candidates for these applications \cite{Drewsen2009} due to their large rotational constants and Doppler-cooled dissociation products.
	
To date, spectroscopy on CaH$^+$  is limited to two vibrational overtones,\cite{Khanyile2015} four vibrational levels within the $2^{1}\Sigma$ state,\cite{Rugango2016} a photodissociative electronic transition,\cite{Hansen2012} and single-ion quantum logic probes of rotational state. \cite{Chou2017}  The vibronic transitions of Ref. \citenum{Rugango2016} were previously assigned according to theory. The observed transition frequencies agreed to within 50 cm$^{-1}$ of theory, but the $v=0 \rightarrow v'=0$ transition was not observed. A similar issue was encountered for the isoelectronic KH neutral, where the first few vibronic lines were experimentally absent and KD spectroscopy was required in order to correctly assign the KH transitions. \cite{Pardo1983, Pardo1987, Stwalley1991, Bartky1966} Isotopic substitution changes the reduced mass but maintains the adiabatic electronic potential energy.  The resulting shift in vibrational energy levels can be compared to theory to confirm peak assignments.
	
Here we apply this method to the spectroscopy of CaH$^+$  and CaD$^+$. Vibronic transitions were calculated using a CASPT2 internuclear potential \cite{Abe2012} and then compared to measured transition frequencies obtained by resonance enhanced multiphoton photodissociation spectroscopy (REMPD). Instead of observing the predicted shifts for deuterium substitution based on our previous assignment, this study reveals a 687 cm$^{-1}$ disagreement in the electronic energy from CASPT2 and leads to a revised labeling of the CaH$^+$  and CaD$^+$ vibronic transitions.

\section{Methods}

\begin{figure}[b]
	%\centering
	\includegraphics[height=6.8cm]{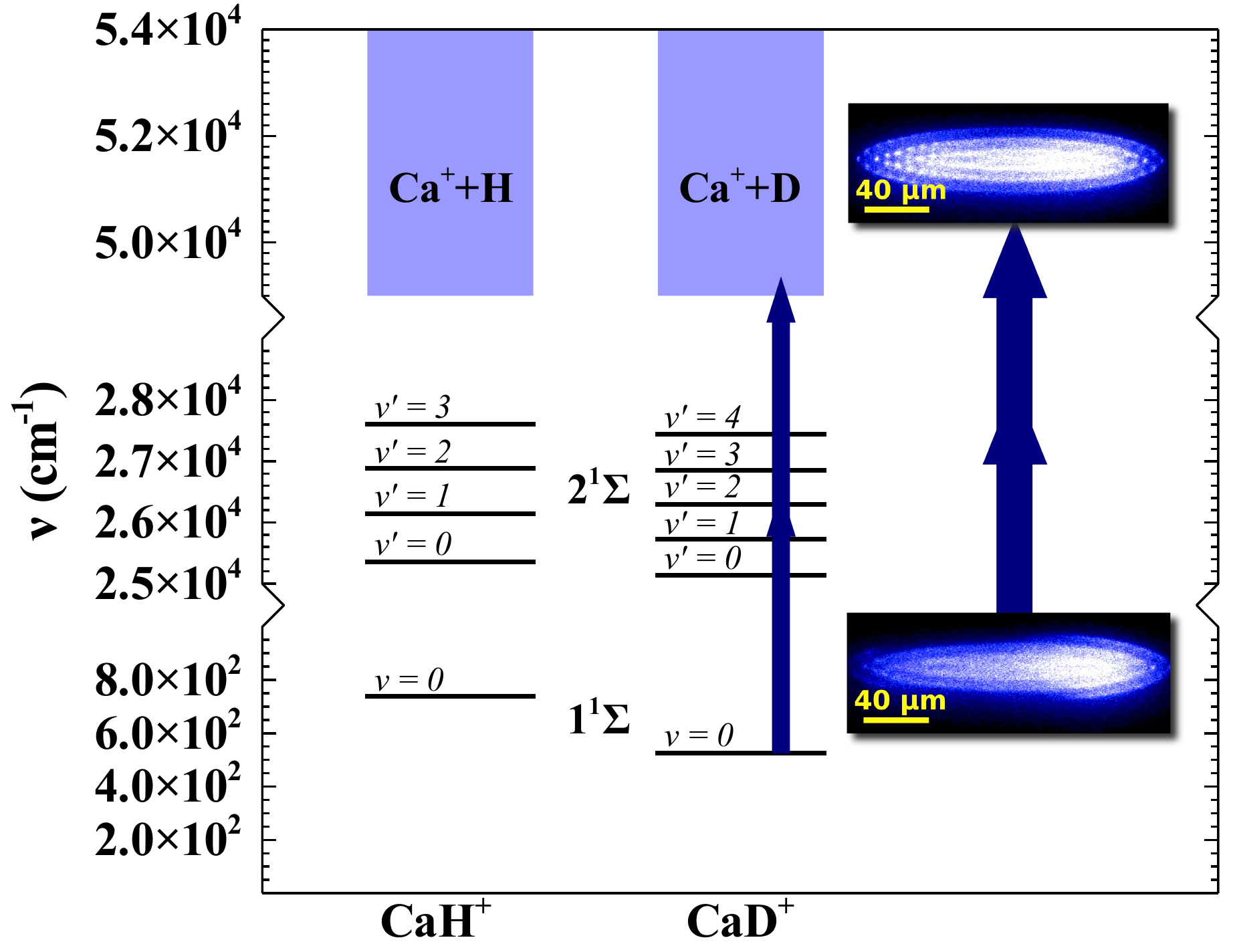}
	\caption{Energy level diagram of the CaH$^+$ and CaD$^+$ transitions probed by applying a doubled Ti:Sapphire laser and observing resonance enhanced photodissociation.  The photodissociation of dark CaD$^+$  ions in the crystal into trapped Ca$^+$ and free {H} can be observed in the crystal image and by an increase in 397 nm Ca$^+$ fluorescence signal.}
	\label{simple}
\end{figure} 

\subsection{Experimental Setup}

The experiment employed a manually-tuned, frequency-doubled Ti:Sapphire laser to probe the vibronic transitions of CaD$^+$  co-trapped with laser-cooled Ca$^+$ in a heterogeneous Coulomb crystal. By measuring the Ca$^+$ fluorescence increase upon laser exposure, a spectrum of molecular dissociation rate was generated.  The experiments take place in an ultrahigh vacuum chamber with a background pressure of 10$^{-10}$ torr and the ions are held in a linear Paul trap.  CaD$^+$ is generated by reaction of D$_2$ with excited Ca$^+$ at pressures of 10$^{-8}$ torr.  The Ca$^+$ is observed and laser cooled by the laser-induced fluorescence at 397 nm. A repump laser at 866 nm is also required to close the transition. Details of the experimental setup may be found in our previous work\cite{Rugango2016} on the vibronic spectroscopy of CaH$^{+}$. 

\begin{figure}[t]
	%\centering
	\includegraphics[height=5.5cm]{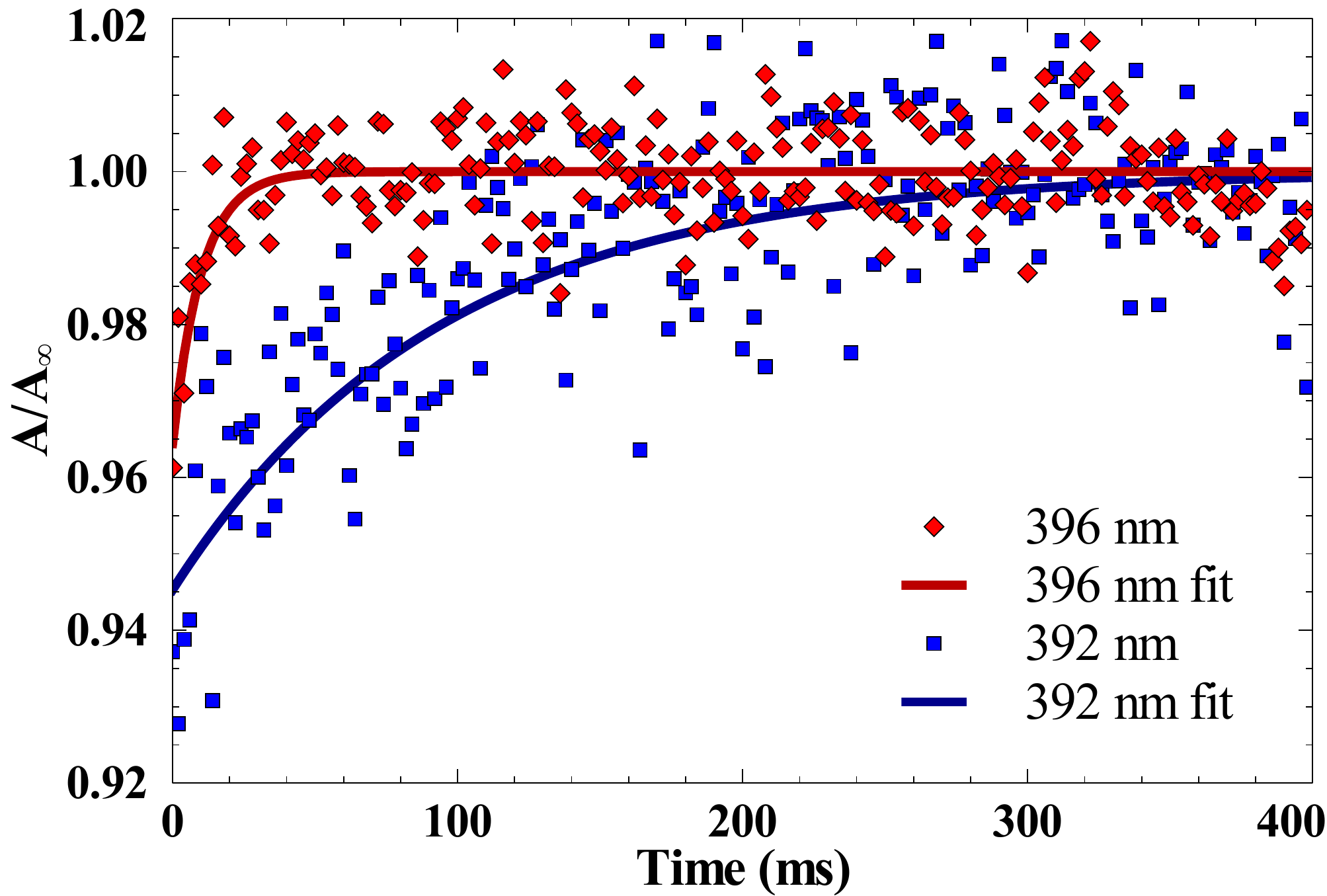}
	\caption{Observed Ca$^+$ fluorescence of composite Coulomb crystals exposed to two laser frequencies. An increase in fluorescence indicates resonance-enhanced dissociation of CaD$^+$  into Ca$^+$ and D. Dissociation rates extracted from Eqn. \ref{eqn:diss} are plotted against frequency in Fig. \ref{fig:comparison}.}
	\label{fig:scans}
\end{figure}

\begin{figure*}[t]
	\centering
	\includegraphics[height=6cm]{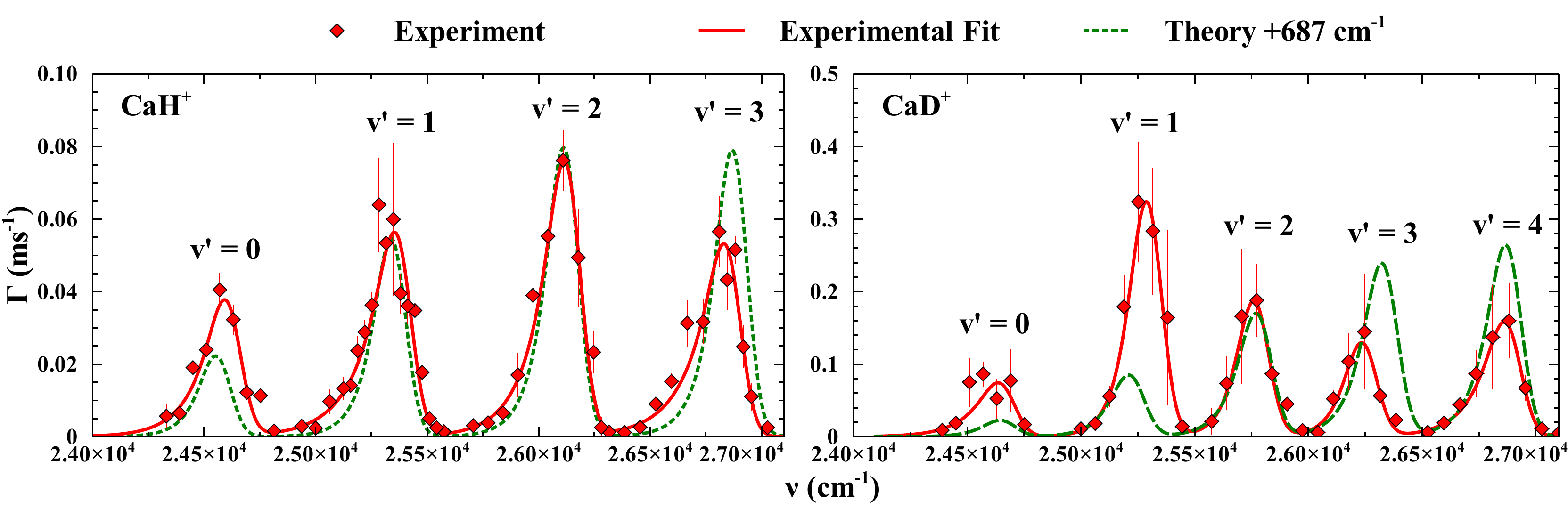}
	\caption{Comparisons of the experimental and \textit{ab initio}-predicted spectra for each isotopologue. The optimized parameters include $G_{1}(v')$, $B_{1,v'}$, and $\mu_{v\rightarrow v'}$. Relative heights of theoretical vibronic peaks are governed by CASPT2 transition dipole moments scaled to fit the experimental $v'=2$ peak, while peak shapes come from the Boltzmann rotational state distribution 298 K.}
	\label{fig:comparison}
\end{figure*}

The CASPT2 potential energy surface of CaH$^+$  guided our spectroscopic search for CaD$^+$  vibronic transitions.\cite{Abe2012} To achieve the desired 24390 to 27100 cm$^{-1}$ range, the mode-locked Ti:Sapphire laser was frequency-doubled by a BBO crystal before being sent along the trap axis. Each fluorescence measurement was taken for 8 ms after ten sets of alternating 200 $\mu$s delay and 200 $\mu$s exposure to the AOM-shuttered 20 mW Ti:Sapph beam. A fit of the fluorescence intensity to the total exposure time \textit{t} to the exponential equation

\begin{equation}\label{eqn:diss}
A_{t}=A_{\infty}-(A_{\infty}-A_{0})e^{-\Gamma (\lambda) t}
\end{equation}
yields the dissociation rate $\Gamma$ as a function of wavelength $\lambda$. $A_{\infty}$ and $A_0$ are the steady-state and initial fluorescence counts, respectively. Scans exhibiting first-order CaD$^+$  dissociation are presented in Fig. \ref{fig:scans}. Dissociation rate plotted as a function of frequency yielded the spectrum in Fig. \ref{fig:comparison}.\footnote{The CaH$^+$ peaks were taken with ten sets of alternating 400 $\mu$s delay and 400 $\mu$s exposure time. Ref. \cite{Rugango2016} peak heights and rates differ by a factor of 10 because an internal counter in the computer control of the experiment was overlooked in the analysis. }

\subsection{Theoretical Model for Parameter Estimation}

To compare theory and experiment, CaD$^+$  dissociation spectra were modeled using theoretical and experimental parameters. Calculation of the Einstein $A$ for each transition gave a dissociation rate, and the total rate at each wavelength was obtained by summing all transitions covered by the laser linewidth. \cite{Rugango2016} Although the experimental CaD$^+$  dissociation is a multi-photon process, we used a first-order model by assuming the dissociation rate is much greater than the first excitation rate. 

The spectra labeled ``Theory" in Fig. \ref{fig:comparison} rely heavily on CASPT2 \textit{ab initio} calculations: the potential energy surfaces, transition dipole moments $\mu_{v\rightarrow v'}$, and vibronic transition frequencies \textit{$G_{0\rightarrow1}(v')$} come from Ref. \citenum{Abe2012}. Harmonic, $\omega_1$, and anharmonic, $\omega\chi_1$, constants were generated using the internuclear potential curves and R.J. LeRoy's open-source project, Level 8.2. \cite{Roy2017} Relative peak heights were determined by assuming a thermal distribution of the rotational states in 1$^{1}\Sigma$. The first 15 \textit{J} rotational levels contribute $>$99.99\% of the initial state population and dominate the shapes of the vibronic peaks. 

During spectral simulation, the spectral density of the laser was held at a constant linewidth of ~80 cm$^{-1}$ to match the laser linewidth based on spectrometer measurements. The model's invariant peak intensity of 4.77 $\times$ 10$^{8}$ W/m$^2$ was calculated from the laser power and beam waist diameter at the trap center. The \textit{ab initio} calculated energy of the $1^{1}\Sigma$ ground vibrational state was used as a reference for all transitions due to lack of experimental data on the ground state potential. 

\section{Results and Discussion}

\begin{table}[b]
	\caption{Observed transitions from 1$^{1}\Sigma (v=0) \rightarrow 2^{1}\Sigma (v=v')$. Frequencies are determined by fitting a convolution of laser linewidth and parameter-dependent level structure to laser-induced dissociation rates of CaH$^+$  and CaD$^+$. The resulting experimental fit is compared to an \textit{ab-initio} spectrum in Figure \ref{fig:comparison}. The frequency in parentheses was unobserved but predicted by the CASPT2 model.}
	%	Extraction of the $B_v$ constants from Ref. \citenum{Abe2012} resulted in mismatched frequencies when renumbering transitions. If  was instead a free parameter during experimental fit generation, the simulation would converge on the same $B_v$ constant, independent of assignment. All values are in cm$^{-1}$.}
	\label{tab:theory_compars}
	{\small
		\hfill{}
		\begin{tabular}{c|c c c}
			\hline
			%$v'$ & \multicolumn{4}{c}{$\nu_{v'}$}\\
			&Previous CaH$^+$ &Revised CaH$^+$ & CaD$^+$\\
			$v'$	&Assign: -50 cm$^{-1}$ &Assign: +687 cm$^{-1}$ &Experimental\\
			\hline		
			
			0 &  (23887) 		& 24635 $\pm$ 49 & 24683 $\pm$ 128 \\
			1 &  24635 $\pm$ 74 & 25401 $\pm$ 21 & 25321 $\pm$ 22 \\
			2 &  25401 $\pm$ 31 & 26158 $\pm$ 18 & 25792 $\pm$ 49 \\
			3 &  26158 $\pm$ 27 & 26879 $\pm$ 29 & 26268 $\pm$ 60 \\
			4 &  26879 $\pm$ 35 & - 			 & 26908 $\pm$ 55 \\
			\hline
	\end{tabular}}
	\hfill{}
	%\caption{Table Name}
	\label{transitions}
\end{table}

\subsection{Comparison of Spectra}

Fig. \ref{fig:comparison} compares the REMPD spectra of CaH$^+$  and CaD$^+$. As expected, the CaD$^+$  transition frequencies were more tightly-packed, owing to the decrease in $\omega_e$ accompanying the increase in reduced mass. Complication arose when matching the \textit{ab initio} predictions to experimental values: the theory, shifted by -50 cm$^{-1}$ to agree with CaH$^+$, systematically underestimated the CaD$^+$  vibronic transition frequencies by $>$100 cm$^{-1}$. If the assignments proposed in Ref. \citenum{Rugango2016} were correct, this study would imply a 150 cm$^{-1}$ isotopic shift of the electronic energy potential and further evidence the experimentally unobserved transition to the $v'=0$ state. The isotopic shift of the electronic energy level is large compared to the 10 cm$^{-1}$ shifts seen in KH.\cite{Pardo1987, Bartky1966} These spectral anomalies prompted a reassignment of the vibrational energy levels within the $2^{1}\Sigma$ manifold.

The new assignment of vibrational quantum numbers, shown in Table \ref{tab:theory_compars}, reflects a 687 cm$^{-1}$ departure from \textit{ab initio} calculations. This shift manifests as a 687 cm$^{-1}$ increase in $T(1)-E_0$. Roughly a vibrational quantum in CaH$^+$, the revision is greater than the 100 - 150 cm$^{-1}$ error window afforded by the mismatch of calculated and measured dissociation asymptotes for Ca$^+$$(3d^{1})$ and H$(1s^{1})$. This revised assignment, however, features observable vibrational peaks through $v'=4$ and good agreement for both isotopologues. 

%Under both numbering schemes, the radial asymptote requires readjustment, so we conclude that the potential of the $2^{1}\Sigma$ state is shallower than initially predicted.

\subsection{Experimental Parameters}

\begin{figure}[!b]
	%\centering
	\includegraphics[height=6cm]{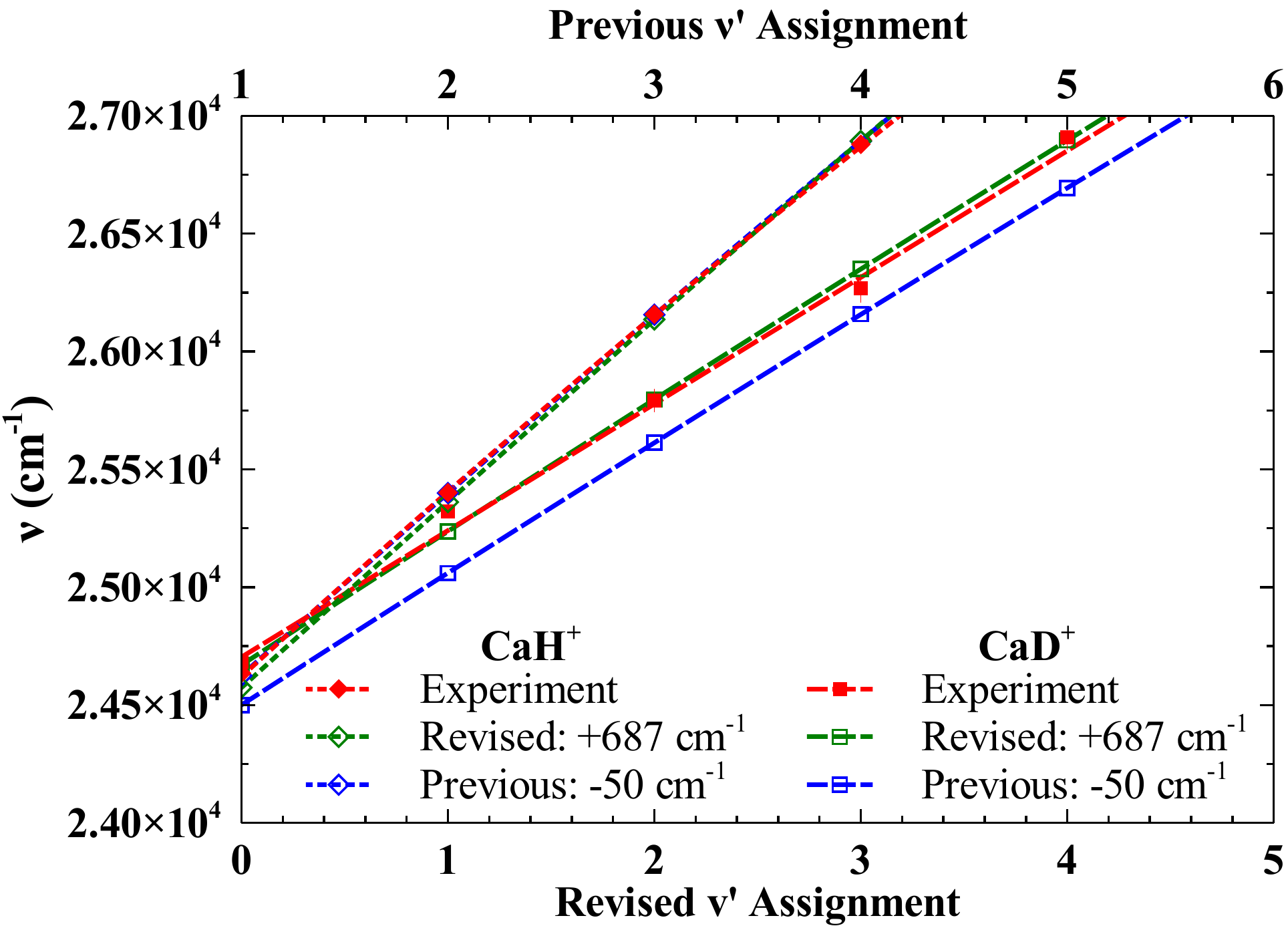}
	\caption{The vibronic transitions of CaH$^+$ and CaD$^+$ and their associated fits using Eqn. \ref{eqn:v}. Theoretical and experimental values are separated by 687 cm$^{-1}$ on average. Spectroscopic constants of each species may be found in Table \ref{tab:constants}.}
	\label{linear}
\end{figure}

With the new CaH$^+$  and CaD$^+$ peak assignments, we determined the spectroscopic constants of the excited state by fitting both theory and experimental values to a second-order model of vibrational energy levels. Vibronic transitions to $v'$ of the $2^{1}\Sigma$ state were modeled with the equation

\begin{equation}\label{eqn:v}
v_{v'} = T(1) + \omega_1 (v'+\frac{1}{2}) - \omega\chi_1(v'+\frac{1}{2})^2 - E_0 
\end{equation}
where \textit{T}(1) is the potential minimum of the 2${^1}\Sigma$ state, and $E_0$ is the zero-point energy of the ground state. The parameters $T(1)$, $\omega_1$, and $\omega\chi_{1}$ for both CaH$^+$  and CaD$^+$ are varied to fit experimental data points by quadratic regression. Regression curves of theory predictions and experimental fits were plotted (see Fig. \ref{linear}) to obtain the constants listed in Table {\ref{tab:constants}}. Since CaH$^+$  ground state information is limited,\cite{Khanyile2015} the $E_0$ energy is confined to the \textit{ab initio} prediction. 

The spectroscopic constants agree reasonably with theory, however we notice a large shift of the excited state potential. To maintain the assumption that the internuclear potentials of CaH$^+$  and CaD$^+$ are similar, the energy surfaces of the \textit{ab initio}-calculated $1^{1}\Sigma$ and $2^{1}\Sigma$ states must be separated by an additional 687 cm$^{-1}$.

This sizable departure from \textit{ab initio} calculations is currently unexplained. Deviations could be due to the frozen core electron approximation of CASPT2 or non-adiabatic effects.  The latter effect is expected to be small based on estimations from other diatomic hydrides. LiH vibrational levels show a 10 cm$^{-1}$ isotopic electronic shift\cite{Zrafi2006, Tung2011} and the KH, KD system also features corrections around 10 cm$^{-1}$.\cite{Pardo1987, Bartky1966} The disparity in experimental and calculated peak heights may be explained after further study of electronic levels relevant to the dissociation pathway. In addition, the different rates for CaH$^{+}$ and CaD$^{+}$ dissociation may be due to an enhanced forward reaction  rate for Ca$^{+} + $ H$_{2}$ due to the kinetic isotope effect depressing the observed dissociation rate. 

In our computations, the 102 cm$^{-1}$ uncertainty in $\omega_1$ absorbs the CaD$^+$ $\omega\chi_1$ constant. The deviations in peak position and relative peak heights from our simple model may be due to an unknown dissociation pathway and is a topic of future research. In particular, the shape of the v'=0 peak was not well matched by our simple model leading to a large uncertainty in its calculated position.

\section{Conclusion}

The vibronic spectrum of CaD$^+$  was obtained by scanning a frequency-doubled Ti:Sapph laser over the frequencies predicted to excite $2^{1}\Sigma$ vibrational modes before coupling to unbound electronic states. Collection of Ca$^+$ fluorescence allowed us to quantify the rate of CaD$^+$  dissociation and plot it against frequency at constant laser linewidth and intensity. The harmonic constant $\omega_{1}=537\pm102$ cm$^{-1}$ and anharmonicity $\omega\chi_{1}=0\pm20$ cm$^{-1}$ of the $2^{1}\Sigma$ state were extracted by fitting our experiments to a second-order vibrational energy expression, Eqn. \ref{eqn:v}. Comparison of the simulated experiment with CASPT2 predictions revealed a 687 cm$^{-1}$ average deviation from standing theory. CaH$^+$  vibrational levels within $2^{1}\Sigma$ were consequently reassigned by reducing the vibrational quantum number by one relative to the previous assignment.\cite{Rugango2016}

\begin{table}[t]
	\small
	\caption{\ Molecular constants for the 1$^{1}\Sigma\rightarrow$ 2$^{1}\Sigma$ vibronic transitions of CaH$^+$ and CaD$^+$ based on the revised peak assignments. All values are in cm$^{-1}$.}
	\label{tab:constants}
	{\hfill
	\begin{tabular}{c|cccc}
		\hline
		& CaH$^+$ & CaH$^+$  &  CaD$^+$&  CaD$^+$\\
		&Experimental &CASPT2 &Experimental &CASPT2 \\
		\hline
		${\omega{_1}}$ & 795 $\pm$ 12 &  803 $\pm$ 3 &537 $\pm$ 102 & 574 $\pm$ 1\\
		${\omega\chi{_1}}$ & 11.5 $\pm$ 3.0 & 8.0 $\pm$ 0.7 & 0 $\pm$ 20 & 3.5 $\pm$ 0.1\\
		$T(1)$ & 24978 $\pm$ 10 & 24226 $\pm$ 2  & 24973 $\pm$ 108& 24223 $\pm$ 1\\
		$E_0$ & - & 739 & - & 526\\
		\hline
	\end{tabular}
	}
\end{table}

To understand this disagreement, we require further information on the ground and excited electronic states of CaH$^+$. A theoretical approach is to recalculate the molecular potentials with unfrozen core electrons. Experiments using mid-infrared spectroscopy to measure lower vibrational transitions of the ground-state can be combined with previous vibrational overtone data \cite{Khanyile2015}  to construct a $1^1\Sigma$ potential energy. Probing higher-lying electronic states could offer insight into photodissociation rates and explain why they differ from theoretical expectations. Future experiments will also include examination of the $2^{1}\Sigma$ state with rotational resolution, improving the precision of the constants presented here. \\

\begin{acknowledgments}
The authors acknowledge funding from the Army Research Office under award W911NF-12-1-0230 and Multi-University Research Initiative award W911NF-14-1-0378. We also acknowledge the National Science Foundation award PHY-1404388. Thanks to Colin Trout for his input on the manuscript. 
\end{acknowledgments}

\end{document}